\documentclass[12pt]{book}

\usepackage[dvips]{graphicx,color}
\usepackage{makeidx,tsukuba}
\makeauthorindex
\makeindex

\begin{document}

\BookTitle{\itshape The 28th International Cosmic Ray Conference}
\CopyRight{\copyright 2003 by Universal Academy Press, Inc.}
\pagenumbering{arabic}

\chapter{VHE $\gamma$-rays From Extragalactic Large Scale Jets}

\author{
\L ukasz Stawarz$^{1, \star}$, Micha\l \ Ostrowski$^1$ and Marek Sikora$^2$ \\
{\it (1) Astronomical Observatory of the Jagiellonian University,
Orla 171, 30-244 Cracow, Poland \ ($^\star$stawarz@oa.uj.edu.pl) \\
(2) N. Copernicus Astronomical Center,
Bartycka 18, 00-716 Warsaw, Poland}}

\section*{Abstract}

We discuss production of very high energy (VHE) $\gamma$-rays in large
scale extragalactic jets of FR I sources. Based on recent optical and X-ray
observations of these objects we evaluate their VHE emission in a framework
of one electron population model. We consider possibility for detection of
the predicted radiation, which could provide important constraints on the jet
structure and some hardly known large scale jet parameters.

\section{Introduction}

Among several discovered sources of VHE $\gamma$-ray emission (photon energies
$\varepsilon_{\gamma} > 100$ GeV), the most numerous group consists of
low-luminosity blazars (BL Lacs). Up to now, there are four confirmed
objects of this kind, including the well known Mrk 421 and Mrk 501,
while three others are classified as probable or possible ones (e.g., [8]).
A widely accepted model explaining production of the TeV emission in blazars
is a synchrotron self-Compton (SSC) process, in which energetic electrons
flowing within a relativistic jet at sub-parsec distances from the active
galactic nucleus (AGN) inverse-Compton upscatter their own synchrotron photons
(e.g., [7]). Although limited to a few cases, TeV blazars attract a great
attention due to at least three reasons. Firstly, by studying $\gamma$-ray
emission of these objects one gets insight into physical conditions in the
closest vicinity of AGN and its nuclear jet. Secondly, TeV blazars are crucial
in understanding acceleration processes acting within the relativistic outflows,
most likely connected with ultrarelativistic shock waves. And finally,
detection of the VHE radiation from extragalactic sources has important
cosmological consequences, because of the photon-photon interaction expected
to occur during propagation of $\gamma$-rays on cosmological distances (e.g.,
[1]). Due to importance of the considered studies, a lot of both theoretical
and observational efforts are directed to identify new BL Lacs radiating in
the VHE range (e.g., [5]).

In the present paper we propose, that an another class of the VHE objects can
be formed by kpc-scale jets of FR I radio galaxies. In particular, we suggest
that TeV signal detected 30 years ago from the Centaurus A system, as well as
the most recent VHE detection from M 87 radio galaxy, can result from the
large-scale jet inverse-Compton (IC) emission.

\section{Methods and Results}

Several jets in FR I radio galaxies are now being observed not only at
radio frequencies, but also at higher (up to X-rays) photon energies.
Synchrotron nature of their radio-to-X-ray radiation seems to be already
established, implying presence of extremely energetic electrons, with
energies $E_e \geq 10$ TeV for the standard equipartition magnetic field
$10^{-4}$ G. Spectral character of the optical and X-ray emission, as well
as its spatial distribution, suggests necessity and complexity of the
particle acceleration processes involved, which in addition seem to be closely
related to the jet internal structure [10]. However, as a rough simplification
one can generalize that in the brightest knots of the discussed objects the
synchrotron continuum is a broken power-law, with a spectral break -- where
most of the synchrotron power is emitted -- placed usually around $10^{14}$
Hz, or, like in the case of M 87, even at higher frequencies. The isotropic
luminosity of the FR I optical/X-ray jets is usually in the range $10^{39} -
10^{42}$ erg/s.

In addition to the synchrotron radio-to-X-ray emission, the FR I jets have
also to produce high energy $\gamma$-rays by IC scattering of ambient photon
fields, including the CMB radiation or the synchrotron emission produced by
the jet itself. At the kpc distances from the galactic center, additional sources
of ambient photons are the AGN and the host galaxy. The AGN radiation
consists of isotropically emitted component connected with thermal gas
and/or dust heated by the central source (narrow line region, dusty nuclear
torus), plus an anisotropic blazar-like emission due to highly relativistic
nuclear jet. A dominating component of the host galaxy radiative output originates
from the stellar and circumstellar dust emission. The considered seed photon fields
differ in intensity, in characteristic photon energies, and in spatial distribution
with respect to the large scale jet. Unfortunately, parameters which are crucial
in determining their relative importance in the jet comoving frame -- like the jet
bulk Lorentz factor $\Gamma$ and its inclination to the line of sight $\theta$ --
are only poorly constrained, or even unknown. For example, radio observations of
FR I sources indicate presence of only moderate or weak relativistic beaming effects.
However, one cannot exclude possibility that in the cases of the optical/X-ray FR I
jets the bulk Lorentz factors are of order of a few. This idea is supported by {\it
Hubble} observations of superluminal motions at large scales in the jet of M87 [3].

For the given break frequency of the jet synchrotron continuum, its luminosity and,
finally, the seed photon fields contributing to the IC scattering, one can estimate
the expected fluxes and photon break energies of the respective $\gamma$-ray emissions.
We provide such estimates in [11] for selected jet models, including possible
relativistic jet bulk velocities. Because of a relatively wide physical range available
for the mentioned parameters of the models, it is difficult to discuss $\gamma$-ray
radiation of the considered objects in general. Instead, a more detailed analysis can
be performed for chosen individual sources. One of our general findings is, however,
that comptonization of the narrow line and the nuclear tours emissions (which are
typically weak or even absent in FR I radio galaxies) in the kpc-scale jets is always
negligible as compared to comptonization of the blazar photons. On the other hand, the
latter process can be significantly suppressed by the Klein-Nishina (KN) effects, if
only a critical frequency of the blazar synchrotron emission is in the optical range,
or higher. In the considered sources we also find that at the kpc distances from the
galactic center the energy density of the stellar emission can be quite large,
exceeding the CMB energy density even by three orders of magnitude. The resulting
IC emission is expected to dominate the jet VHE radiative output for small or moderate
jet inclinations and relativistic bulk velocities, limited only by the KN effects.
The SSC process can play more significant role for larger jet inclinations.

\section{Discussion: Centaurus A and M 87}

In [11] we discuss $\gamma$-ray emission of Centaurus A and M 87 large scale jets.
Both sources are relatively nearby, what allowed in the past to study their jets
at different frequencies and scales, with high spatial and spectral resolutions.
During the last decades, M 87 and Centaurus A were also frequently observed at
$\gamma$-rays, from sub-MeV to TeV photon energies, what resulted in positive
detections or constraints on the flux upper limit.

{\it CANGAROO} observations of the extended ($\sim 14$ kpc) region around the
Centaurus A nucleus put the upper limit on the emission at the VHE range,
$S(\varepsilon_{\gamma} \geq 1.5 \, {\rm TeV}) < 1.28 \cdot 10^{-11} \,
{\rm ph \, cm^{-2} \, s^{-1}}$ [9]. However, the discussed object was detected
in the past at TeV energies, with the observed flux $S(\varepsilon_{\gamma} \geq
0.3 \, {\rm TeV}) \sim 4.4 \cdot 10^{-11} \, {\rm ph \, cm^{-2} \, s^{-1}}$ [6],
what made Centaurus A the very first extragalactic source of the detected VHE
radiation. In a framework of our model, for the parameters usually considered for the
large scale jet in this source -- i.e. equipartition magnetic field, large jet
viewing angle $\theta \sim 70^0$ and $\Gamma \leq 2$ -- the SSC radiation seems to
be able to overproduce (comparing to the present upper limits) the VHE photons.
This suggest presence of the KN effects decreasing the SSC emission, what in turn
indirectly suggests relativistic jet velocities and/or subequipartition magnetic
field. We also find, that for the moderate and large jet inclinations, comptonization
of the nuclear blazar-like emission of Centaurus A (see, e.g., [4]) is expected to
be very strong, with the observed flux $S(\varepsilon_{\gamma} \geq 10 \, {\rm GeV})
\sim 10^{-11} - 10^{-10} \, {\rm erg \, cm^{-2} \, s^{-1}}$ (for $\Gamma = 1 - 10$).
This emission could be therefore easily observed in the future by {\it GLAST}.

Recent {\it HEGRA} observations resulted in a positive detection of the VHE emission
from the M 87 radio galaxy, at the level $S(\varepsilon_{\gamma} \geq 0.73 \,
{\rm TeV}) \sim 0.92 \cdot 10^{-12} \, {\rm ph \, cm^{-2} \, s^{-1}}$ [2]. Our analysis
suggests, that if this TeV flux is in fact due to the IC large scale jet emission, its
most likely origin is comptonization of the galactic stellar and circumstellar dust
radiation, consistently with moderate jet inclinations $\theta \sim 20^0 - 30^0$ and
relativistic jet velocities $\Gamma > 2$ inferred from the {\it Hubble} observations.
However, because of the expected KN effects, a detailed numerical analysis is required
in order to provide constraints on the jet bulk Lorentz factor and its magnetic field.

\section{Conclusions}

Detection of the VHE $\gamma$-ray emission from, at least, some of the FR I
optical/X-ray jets is possible. Future observations (or, alternatively, non
detections) by ground-base imaging atmospheric Cherenkov telescopes and {\it
GLAST} will therefore provide important constraints on the unknown jet parameters,
like the magnetic fields intensity and the jet kinematic factors. Our analysis
enables to evaluate high energy $\gamma$-ray emissions of the nearby FR I sources
Centaurus A and M 87 [11]. For Centaurus A we predict measurable -- by future
$\gamma$-ray missions -- fluxes at $\varepsilon_{\gamma} \sim 10$ GeV and at
$\varepsilon_{\gamma} \sim 0.1 - 1$ TeV photon energies due to comptonization of
the blazar radiation and the SSC process, respectively. In the case of M 87 we
proposed, that recently detected VHE emission resulted form the IC scattering of
the stellar and circumstellar infrared photons of the host galaxy.

\section{Acknowledgments}

The present work was supported by KBN through the grant
PBZ-KBN-054/P03/2001, by the Department of Energy contract to SLAC no.
DE-AC3-76SF00515, and by NASA Chandra grants via SAO grant no. GO1-2113X.

\section{References}

\re
1.\ Aharonian F.\ 2001, in Proc. of the 27th ICRC (astro-ph/0112314)
\re
2.\ Aharonian F.\ et al.\ 2003, A\&A, 403, 1
\re
3.\ Biretta J.A., Sparks W.B., Macchetto F.\ 1999, ApJ, 520, 621
\re
4.\ Chiaberge M., Capetti A., Celotti A.\ 2001, MNRAS 324, 33
\re
5.\ Costamante L., Ghisellini G.\ 2002, A\&A 384, 56
\re
6.\ Grindlay J.E.\ et al.\ 1975, ApJ 197, 9
\re
7.\ Kino M., Takahara F., Kusunose M.\ 2002, ApJ 564, 97
\re
8.\ Ong R.A.\ 2003, `The Status of VHE Astronomy' (astro-ph/0304336)
\re
9.\ Rowell G.P.\ et al. 1999, APh 11, 217
\re
10.\ Stawarz \L ., Ostrowski M.\ 2002, ApJ 578, 763
\re
11.\ Stawarz \L ., Sikora M., Ostrowski M.\ 2003, ApJ, {\it submitted} (astro-ph/0306251)

\endofpaper
\end{document}